\documentclass[preprint,12pt]{aastex}
\usepackage{epsfig}
\begin{document} 

\title{Discovery and Study of Nearby Habitable Planets with Mesolensing} 
\author{Rosanne Di\thinspace Stefano \& Christopher Night}
\affil{Harvard-Smithsonian Center for Astrophysics, 60
Garden Street, Cambridge, MA 02138}

\begin{abstract}
We demonstrate that gravitational lensing can be used to discover and
study planets in the habitable zones of nearby dwarf stars.  If
appropriate software is developed, a new generation of monitoring
programs will automatically conduct a census of nearby planets in the
habitable zones of dwarf stars. In addition, individual nearby dwarf
stars can produce lensing events at predictable times; careful
monitoring of these events can discover any planets located in the
zone of habitability.  Because lensing can discover planets (1) in
face-on orbits, and (2) in orbit around the dimmest stars, lensing
techniques will provide complementary information to that gleaned
through Doppler and/or transit investigations.  The ultimate result
will be a comprehensive understanding of the variety of systems with
conditions similar to those that gave rise to life on Earth.
\end{abstract}

\section{Introduction: Lensing by Planets in the Habitable Zone}
Microlensing has been considered as a method best suited to discover
distant planets in, e.g., in the Magellanic Clouds, M31, the Galactic
Bulge, Halo, and several kiloparsecs from us in the Galatic
disk. Recently, it has been pointed out that lensing can provide a way
to discover and study nearby (within $\approx$ 1 kpc) planetary
systems, providing complementary information to that obtained from
Doppler and/or transit studies \citep{DiStefano2007}.  Nearby planets
are intriguing because detailed observations are possible.  If life
exists on other planets, we will find it first and learn most about it
on planets within a few hundred parsecs.

The importance of water to Earth life has suggested that we define a
zone of habitability around each star to be the region in which liquid
water can exist on the surface of a planet that has an atmosphere. The
literature suggests that we can express $a,$ the separation between
the central star of mass $M$, and a planet in the habitable zone as

\begin{equation} 
a = 0.13\, {\rm AU}\, h\, \Big({{M}\over{0.25\, {\rm M_\odot}}}\Big)^{{3}\over{2}},  
\end{equation}

\noindent
where $h$ is approximately equal to unity in the middle of the
habitable zone, and to $2/3, 4/3$ at its inner and outer edge,
respectively \citep{Charbonneau2007, Tarter2007, Scalo2007}. There is
presently significant uncertainty in this equation. The nearby dwarf
Gliese 581, with a stellar mass of $0.31\, {\rm M_\odot},$ has three known
planets, one of which has an $h$ value of 1.39 and is generally
considered to be possibly habitable. Another of its planets has an $h$
value of 0.41, and is generally considered not to be habitable
\citep{vonBloh2007, Selsis2007}, though mitigating effects have been
proposed \citep{Chylek2007}.

The idea motivating our work is to explore the action as a
gravitational lens of a nearby dwarf star that hosts a planet in its
zone of habitability. The deflection of light from a distant star by
an intervening mass causes the image of the source to be split,
distorted, and magnified.  When the lens is located in our Galaxy, and
has mass comparable to that of a star or planet, the images of the
source are separated by too small an angle to resolve.  We can,
however, detect the increase in the amount of light received from the
source when its position, $u,$ projected onto the lens plane is
comparable to the Einstein radius, $R_E$.

\begin{equation} 
R_E = 0.13\, {\rm AU}\, \Bigg[\Big({{M}\over{0.25\, {\rm M_\odot}}}\Big)\,  
                             \Big({{D_L}\over{10\, {\rm pc}}}\Big) 
                             \Big(1-{{D_L}\over{D_S}}\Big)\Bigg]^{{1}\over{2}}, 
\end{equation}

\noindent
where $D_L$ is the distance to the lens and $D_S$ is the distance to
the source star. We have assumed that the mass of the planet is a
small fraction of the mass of the star. Expressing $u$ in units of
$R_E,$ the magnification is $34\%, 6\%, 2\%, 1\%$ when $u = 1, 2, 3,
3.5,$ respectively.

\bigskip 

Comparing equations (1) and (2) reveals that the spatial scale that
defines the habitable zone is compatible with the size of the
Einstein ring for a wide range of stellar masses and distances, $D_L.$
This is fortuitous, because the chances of detecting a planet around a
lens star are greatest when the orbital distance is comparable in size
to the Einstein radius.

In this Letter we focus on planetary systems located within about a
kiloparsec. These systems are near enought that we can hope to conduct
follow-up observations to learn more about the planet and perhaps
eventually test for the presence of life.  It is convenient to define
the parameter $\alpha$ to be the ratio between the semimajor axis and
the value of the Einstein radius: $a = \alpha\, R_E.$ Figure 1 shows
values of $\alpha$ as a function of distance, for planets in the zones
of habitability around their stars.  Depending on the stellar mass,
planets in the habitable zone are detectable through their action as
lenses for planetary systems as close to us as a parsec (for the
lowest-mass dwarf stars), and out to distances as far as a kiloparsec
or more when the stellar mass is $0.5\, {\rm M_\odot}$ or larger.

\section{Signatures of Lensing by Planets in the Habitable Zone}

\subsection{Orbital Motion}

Signatures of the lensing by a planetary system have been studied
theoretically \citep{MaoPaczynski, GouldLoeb, DiStefanoScalzo1,
DiStefanoScalzo2, GriestSafizadeh}. Four planets have been observed
\citep{OGLE-2003-BLG-235, OGLE-2005-BLG-071, OGLE-2005-BLG-169}. These
discoveries were made by searching for a signature in which an ongoing
event caused by a star's action as a lens is punctuated by a
short-lived light curve feature caused by the presence of a
planet.

In many cases, nearby planets in the habitable zone will produce
signatures that are similar to those already observed for more distant
systems. Figure 1 shows, however, that the ratio between the event
duration (usually one to three times the Einstein diameter crossing
time, $\tau_E$) and orbital period can be close to or even greater
than unity for these systems. This can produce new types of events,
increase the overall event rate, and decrease some event durations.
Figure 2 illustrates the region of deviation whose geometry and
rotation produce detectable events.

We have conducted a set of simulations for each of a sequence of
nearby systems with planets in the zone of habitability. The systems
we worked with are marked by a cross in Figure 1; the stellar mass is
$0.25\, {\rm M_\odot},$ the orbital separation is $0.1 {\rm AU}$, and
the systems differ from each other only in their distance from us,
which ranges from $10$ pc to $200$ pc.  We have computed the rate of
deviations, for each of several levels of detectability. To
demonstrate that rotation is an important factor for these systems, we
have run the simulation twice: once taking rotation into account and
once assuming no rotation. (See Table 1.) Typical examples of the
deviations produced by our simulation are shown in Figure 3.

\subsection{Detectability}

Several trends are evident in Table 1. Naturally, in all cases, the
greater the sensitivity to deviations, either shorter in
duration or smaller in strength, the larger the rate. In all cases,
caustic crossings are negligible. Generally, the farther that $\alpha$
is from the so-called critical zone around $\alpha = 1$, the fewer the
deviations. The effect is allayed in many circumstances, however, by
rotation. For $\Delta$ = 0.003 and 0.01, we still see a very
appreciable rate of one-hour events even down to $\alpha$ = 0.223.
Because rotation significantly enhances detectability, it is important
to take its contribution into account.

It must be noted that our simulation assumes that the size of the
source star's disk is infinitesimal, an assumption that is justified
in many but not all situations for nearby lenses. Generally, if the
timescale of the deviation $t_{dev}$ is much greater than the time it
takes the lens to traverse the source's disk, finite source size
effects can be neglected. This holds when:

\begin{equation}
t_{dev} \gg 0.04\, {\rm hr}\, \Bigg(\frac{R_*}{\rm R_\odot}\Bigg) 
                              \Bigg(\frac{50\, {\rm km/s}}{v}\Bigg)
                              \Bigg(\frac{D_L}{100\, {\rm pc}}\Bigg)
                              \Bigg(\frac{10\, {\rm kpc}}{D_S}\Bigg)
%t_{dev} \gg \frac{R_*}{v} \frac{D_L}{D_S}
\end{equation}

\section{Expected Event Rate}
\subsection{Targeting a Single Dwarf Star for Lensing Studies} 
Unlike other methods of planet detection, gravitational lensing relies
on light from a more distant star. It is therefore important to ask
what fraction of nearby dwarfs will pass in front of bright sources
and so can be studied with lensing.  Within $50$ pc, there are
approximately $2$ dwarf stars, primarily M dwarfs, per square
degree. Consider a star with $M=0.3\, {\rm M_\odot},$ located at $50$
pc. For this star, $\theta_E = R_E/D_L = 0.006'',$ and, if its
transverse velocity, $v,$ is $50$ km s$^{-1}$, it traverses $0.21''$
yr$^{-1}$. If the path of a more distant source star must cross within
$\theta_E$ of the lens in order for there to be a detectable event,
the lens can generate one event per (year, decade, century) if the
density of the source field (per sq.\,arcsecond) is (16, 5.0, 1.6).
An accurate calculation of the rate of detectable events must consider
that, when we monitor dense source fields, even the smallest
resolution element, $\theta_{mon}$ is likely to contain more than one
star.  In order for the magnification of a particular source star to
be detectable, the increase in the amount of light we receive from it
must be enough to produce a certain fractional change, $f_T$, in the
amount of light received from a region $\theta_{mon} \times
\theta_{mon}$.  The rate at which a single star generates detectable
events \citep{DiStefano2005} is

\begin{equation}
{\cal R}_1 ={{0.027}\over{\rm yr}}\, \Big({{0.1}\over{f_T}}\Big)\,
                           \Big({{1''}\over{\theta_{mon}}}\Big)^2
                           \Big({{v}\over{50 {\rm km/s}}}\Big)\,
                           \Big({{M}\over{0.3\, {\rm M_\odot}}}\Big)^{{1}\over{2}}
                           \Big({{50\, {\rm pc}}\over{D_L}}\Big)^{{3}\over{2}}.
\end{equation}

This rate is high enough that, given a dense background field and a
known nearby M dwarf in front of it, we can plan observations to
detect the action of the dwarf as a lens. For example, if the
photometry can be sensitive enough to allow $f_T= 0.01,$ and if the
angular resolution is good enough to allow $\theta_{mon} = 0.5'',$ the
rate of detectable events can be comparable to one per year, higher if
$f_T$ can be made smaller. Because it is possible for the rate per
lens to be so large, we call nearby lenses ``high-probability lenses''
or {\it mesolenses} \citep{DiStefano2005}.\footnote{The prefix
``meso'' refers to the fact that both spatial and temporal signatures
can be important for nearby lenses.  In cosmological applications of
lensing, spatial signatures can be important, while microlensing is
studied in the time domain.}

This suggests that, by studying the backgrounds behind dwarf stars, we
can predict the times of future events and plan frequent, sensitive,
high-resolution observations to test for evidence of the planet's
effect as a lens \citep{DiStefano2005}.  What can we learn from
targeted observations?  We can determine if there is a planet in the
habitable zone: we will either discover evidence of it, or place a
quantifiable limit on the existence of one. If there is a planet, we
can, under ideal circumstances, measure its mass and its projected
distance from the star.  To plan such observations, we must identify a
large reservoir of potential lenses lying in front of dense
backgrounds. In front of the Magellanic Clouds, e.g., we expect there
to be approximately $200$ ($1600$) dwarf stars with $D_L < 50$ pc
($D_L < 200$ pc).  Equation (4) indicates that, if observations with
small values of $f_T$ and/or $\theta_{mon}$ can be conducted, roughly
$10\%$ of these stars will produce detectable events within a
decade. The first step, therefore, is to study the backgrounds behind
the known dwarf stars lying in front of the chosen background field,
to determine which are the ones most likely to produce events in the
near future, and to predict the likely event times so that appropriate
monitoring observations can be planned for the duration of the
predicted event. For dwarf stars that do not lie in front of dense
fields, serendipitous events are nevertheless possible; one such event
has been observed \citep{Gaudi2007, Fukui2007}. Cross-correlation
between the positions of nearby dwarf stars and the poistions of more
distant sources of light may identify future positional coincidences
that can be predicted with high accuracy \citep{SalimGould}.

\subsection{Monitoring Studies} 
Targeted lensing observations have been suggested by a variety of
authors \citep{Feibelman1986, PaczynskiNearby, SalimGould,
DiStefano2005}, but have not yet been carried out.  Instead,
astronomers have conducted large observing programs in which they have
monitored tens of millions of stars per night \citep{MACHO5.7,
OGLEIIIcatalog, EROSII3year}. More than $3,500$ microlensing event
candidates have been discovered to date. The microlenses tend to be
located at distances greater than several kiloparsecs, and their
presence is revealed through their action as lenses.  While
theoretical work has explored the discovery of nearby lenses by these
programs \citep{DiStefano2007}, several examples of nearby lenses
suggest that dwarf stars constitute as many as $10-20\%$ of the lenses
producing detectable events \citep{Nguyen2004, Kallivayalil2006,
Gaudi2007, Fukui2007}.  For monitoring programs, the expected rate of
events caused by nearby dwarfs, with constant spatial density $N_L,$
in front of a source field of area $\Omega,$ is

\begin{equation}
{\cal R}_{tot} ={{0.069}\over{\rm yr}}\,
                \Big({{\Omega}\over{{\rm sq.\,deg.}}}\Big)
                           \Big({{0.1}\over{f_T}}\Big)\,
                           \Big({{1''}\over{\theta_{mon}}}\Big)^2
                   \Big({{N_L}\over{0.1\, {\rm pc}^{-3}}}\Big)\,
                           \Big({{v}\over{50 {\rm km/s}}}\Big)\,
                           \Big({{M}\over{0.3\, {\rm M_\odot}}}\Big)^{{1}\over{2}}
                           \Big({{D_L}\over{50\, {\rm pc}}}\Big)^{{3}\over{2}}.
\end{equation}

This implies that events must have already been detected by ongoing
monitoring programs, such as OGLE, which has $f_T$ and $\theta_{mon}$
roughly equal to $0.1$ and $1''$, respectively. OGLE monitors almost
100 sq.\,degrees in the Galactic Bulge, but not all at the same
cadence.  Considering just $50$ sq.\,degrees, OGLE data must include
$3.4$ events per year caused by dwarf stars within $50$ pc, and
perhaps as many as $38$ per year caused by dwarf stars within $250$
pc.  Within the next year, a sensitive all-sky monitoring program,
Pan-STARRS will begin monitoring the sky from Hawaii; approximately
$5$ years after that, LSST will begin even more sensitive and frequent
observations from Chile. If values of $f_T = 0.02,$ and
$\theta_{mon}=0.5''$ can be achieved over $200$ sq.\,degrees
(including dense background fields and serendipitous positions of
background stars behind foreground dwarfs), then these programs will
detect events caused by $270$ ($3000$) dwarf stars within $50$ pc
($250$ pc). Each of these events provides an opportunity to test for
the presence of a planet in the zone of habitability.

For monitoring programs, it is worthwhile to consider which signal
will provide a trigger to let us know that a planet-lens event is
occurring. For the now-standard case of microlensing planet detection,
the signature of the planet is a deviation of an ongoing event from
the point-lens light-curve shape. As Figure 3 shows, this can also
happen for planets in the habitable zone when the underlying event has
a large value of $A_{max}$, i.e., when the source track passes close
to the star. In many cases, however, the value of $A_{max}$ is
small. It is therefore important to note that, whether the value of
$A_{max}$ is large or small, the presence of the dwarf star is itself
an important signature. When any deviation is detected, and when there
is a dwarf star along the line of sight, frequent sensitive monitoring
is called for. Such monitoring will either discover evidence of a
planet, or will place limits on the existence of one.

\section{Conclusion}

Until now, searches for habitable planets have relied on very
successful techniques based on Doppler shifts and on transits.  We
point out that gravitational lensing is a potentially powerful
complement to these proven methods. Because it relies on light from
background sources, gravitational lensing can discover planets in the
habitable zones of even the dimmest stars, and can be successful even
if spectral lines and/or the measured flux exhibit erratic variations.
Because it relies only on the distribution of the lens-system's mass,
even face-on systems, inaccessible to Doppler and transit techniques,
can be studied.

The challenge faced by lensing studies is that the lens must pass in
front of a background source of light. This requires that the
planetary system lie in front of a dense source field, or else that
its path happens to bring its projected position close to a background
field star. Predictions \citep{DiStefano2005}, bolstered by the
discovery of events caused by nearby lenses
\citep{Nguyen2004,Kallivayalil2006}, indicate that both types of
events are common.

Monitoring observations of the present and near future, including
OGLE, Pan-STARRS, and LSST can discover events caused by planets in
habitable zones of nearby stars. In most cases, the presence of the
star and its proper motion will already have been catalogued by the
same monitoring data. Software must be developed to recognize even
subtle lensing-like deviations in real time, so that when they occur
at the location of a known possible lens, a separate program of
frequent follow-up observations can be initiated.

Targeted lensing observations are a new frontier. As has been the case
for Doppler and transit studies, it will take time for them to yield
results, but the results are well worth it. For any given nearby dwarf
star, we can discover the presence of a planet, or place limits on the
existence of one. If a planet is discovered, the quantities that can
be derived under ideal circumstances are the planet's mass, its
separation from the star, and its orbital period.

\bigskip
\noindent The authors are grateful to Charles Alcock, David Charbonneau,
Matthew Holman, Penny Sackett, Dimitar Sasselov for conversations.
This work was supported in part by AST-0708924.

\begin{table}[t]
{\scriptsize
\begin{tabular}{ccc|cccc|cccc|cccc|c}
& & &
\multicolumn{4}{c|}{$\Delta = 0.003$} &
\multicolumn{4}{c|}{$\Delta = 0.01$} &
\multicolumn{4}{c|}{$\Delta = 0.03$} &
\\
\hline
$D_L$ & $\alpha$ & $\tau_E/P$ &
$R_0$ & $R_1$ & $R_6$ & $R_{24}$ &
$R_0$ & $R_1$ & $R_6$ & $R_{24}$ &
$R_0$ & $R_1$ & $R_6$ & $R_{24}$ &
$R_{cc}$
\\
\hline
10 & 0.701 & 0.428 &
0.80 & 0.79 & 0.78 & 0.64 & 0.58 & 0.59 & 0.57 & 0.44 & 0.45 &
0.45 & 0.42 & 0.15 & 0.01
\\
10 & 0.701 & 0.428 &
0.68 & 0.68 & 0.68 & 0.64 & 0.52 & 0.51 & 0.51 & 0.44 & 0.41 &
0.41 & 0.40 & 0.22 & 0.01
\\
\hline
20 & 0.496 & 0.604 &
1.48 & 1.50 & 1.48 & 0.96 & 0.90 & 0.89 & 0.79 & 0.22 & 0.37 &
0.37 & 0.29 & 0.04 & 0.001
\\
20 & 0.496 & 0.604 &
0.80 & 0.80 & 0.80 & 0.77 & 0.43 & 0.43 & 0.42 & 0.32 & 0.23 &
0.23 & 0.21 & 0.07 & 0.001
\\
\hline
50 & 0.314 & 0.954 &
1.50 & 1.52 & 1.13 & 0.20 & 0.56 & 0.52 & 0.09 & 0.06 & 0.29 &
0.21 & 0.05 & 0.01 & 0.000
\\
50 & 0.314 & 0.954 &
0.40 & 0.40 & 0.40 & 0.36 & 0.16 & 0.17 & 0.16 & 0.06 & 0.08 &
0.09 & 0.07 & 0.01 & 0.000
\\
\hline
100 & 0.223 & 1.345 &
1.03 & 1.00 & 0.12 & 0.08 & 0.52 & 0.30 & 0.05 & 0.03 & 0.24 &
0.07 & 0.03 & 0.00 & 0.000
\\
100 & 0.223 & 1.345 &
0.19 & 0.19 & 0.18 & 0.13 & 0.09 & 0.09 & 0.08 & 0.02 & 0.04 &
0.04 & 0.03 & 0.00 & 0.000
\\
\hline
200 & 0.159 & 1.891 &
0.96 & 0.26 & 0.06 & 0.04 & 0.52 & 0.06 & 0.03 & 0.01 & 0.09 &
0.02 & 0.02 & 0.00 & 0.000
\\
200 & 0.159 & 1.891 &
0.10 & 0.08 & 0.07 & 0.04 & 0.05 & 0.04 & 0.03 & 0.01 & 0.02 &
0.02 & 0.01 & 0.00 & 0.000
\\
\end{tabular}
}
\caption{ Rates for deviations of various strengths, as determined by
simulation. For $M = 0.25 {\rm M_\odot}$, $M_{PL} = 0.26 M_{Jupiter}$,
and $a = 0.1AU$ held fixed, a face-on system was placed at five
different distances (as shown on Figure 1). For each value of $D_L$
there are two rows, the top including rotation, and the bottom for
(unrealistic) static systems.  Rates are normalized to rate of events
for which $A_{max} > 1.34$. If the number in the table is 1.00, then
there will be as many events with deviations at the given level as
there are events with a maximum magnification greater than 1.34. The
rate of detectable deviations depends on two factors: the size of
deviation $\Delta$ needed for detection, and the duration. For a given
$\Delta$, we define $R_k$ to be the rate of deviations lasting at
least $k$ hours. For instance, if deviations must have a strength of
$1\%$ and last 6 hours to be detectable, their rate may be found in
the $R_6$ column under $\Delta = 0.01$. Although the rotation period
is the same ($P = 23.1$ days) for all systems considered here, each
value of $D_L$ corresponds to a different Einstein radius $R_E$ and
thus a different event duration $\tau_E$. Therefore rotation is
expected to play a larger role for more distant systems. This affects
the event rate by slightly reducing the rate of long-lived deviations,
but greatly increasing the rate of short-lived deviations.  $R_0$ is
the theoretical maximum rate of deviations, if infinitely fine
sampling could be achieved. Values of $R_0$ were determined not by
simulation but by semi-analytic calculation. They should correspond to
the maximum possible rate of deviations, but may be slightly lower
than $R_1$ due to rounding errors.  }
\end{table}

\begin{figure}
\psfig{file=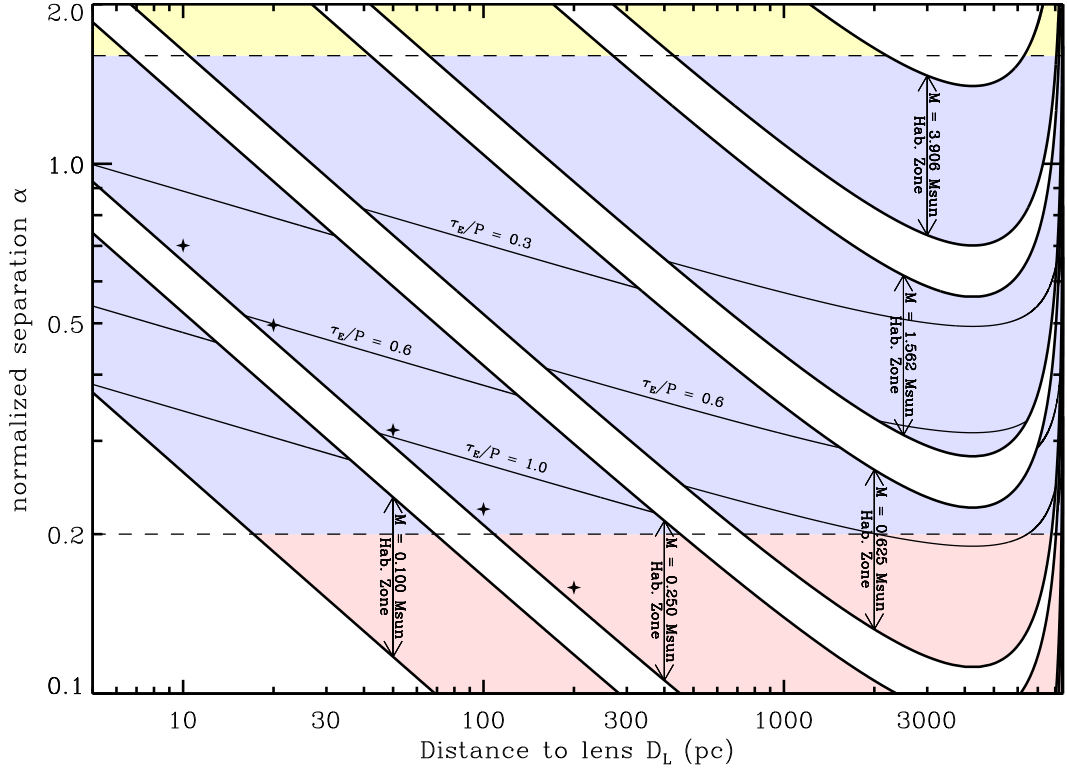,height=11cm}
\caption{ $\alpha$ vs $D_L$ for the habitable zone of low-mass stars.
Each colored bar represents a star with a given mass: $M = 0.1\,
{\rm M_\odot}$ on the lower left, increasing by a factor of 2.5 for each
subsequent bar.  The lower (upper) part of each bar corresponds to the
inner (outer) edge of the habitable zone for a star of that mass.  The
upper horizontal dashed line at $\alpha = 1.6$ marks the approximate
boundary between ``wide'' systems, in which the planet and star act as
independent lenses \citep{DiStefanoScalzo1, DiStefanoScalzo2}, and
``close'' systems in which distinctive non-linear effects, such as
caustic crossings provide evidence of the planet \citep{MaoPaczynski,
GouldLoeb}. All of the planets detected so far have model fits with
$\alpha$ lying between $0.7$ and $1.6.$ In this range, the effects of
caustics are the most pronounced. As $\alpha$ decreases, the effect
of the planet on the lensing light curve becomes more difficult to
discern; the horizontal dashed line at $\alpha = 0.2$ is an estimate
of a lower limit. The probability of detecting the planet in such
close systems ($\alpha \le 0.5$) is increased by the orbital motion. }
\end{figure}

\begin{figure}
\begin{center}
\epsfig{file=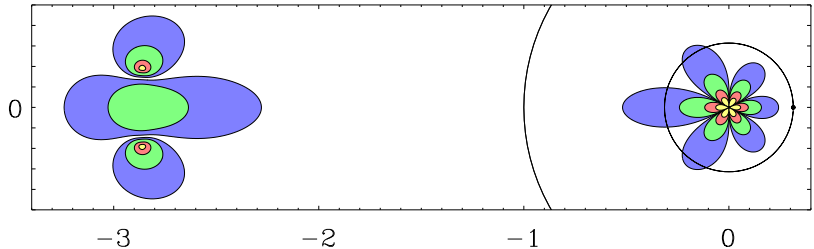,width=14cm}
\end{center}
\caption{ Regions in the lens plane that correspond to deviations from
the point-lens signature. All distance are measured in units of the
Einstein radius. This lens has $M_{planet}/M = 0.001$ and separation
$\alpha$ = 0.314. If the star is 50pc away and has a mass of 0.25
${\rm M_\odot}$, the planet would be in the habitable zone, at an
orbital distance of 0.1 AU. The center of mass is placed at the
origin, and the star is very slightly offset from this. The inner
circle shows the planet's orbit.  At the instant of time depicted, the
planet is located to the right of the star. The outer circle shows the
Einstein radius; if the source passes through this circle, the peak
magnification of the event will exceed 1.34. When the source is within
a colored region, the difference $\Delta$ between the actual
magnification and the expected point-lens magnification exceeds a
certain amount.  Blue (outermost): $\Delta > 0.1\%$.  Green: $\Delta >
0.3\%$.  Red: $\Delta > 1\%$.  Yellow (innermost): $\Delta >
3\%$. Note that the sizes of all of these regions are much larger than
the caustic structures, which are too small to be shown on this
plot. The regions of deviation rotate around the center of mass as the
planet orbits. Since the outer region is much farther from the center,
it moves much faster than the planet does. This effect increases the
probability that the source will pass through a region in which
$\Delta$ is detectable. }
\end{figure}

\begin{figure}
\psfig{file=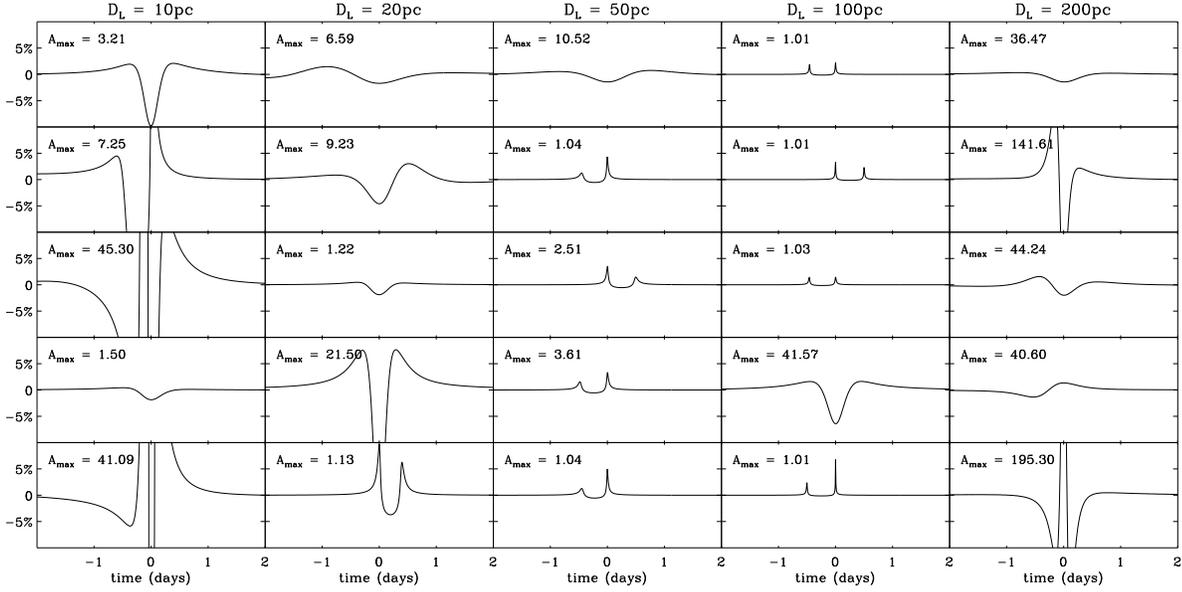,height=7cm}
\vspace{.5 true in} 
\caption{ Typical deviations from the point-lens form, as found with
our simulation. Plotted in each panel is the residual magnification
$\Delta(t)$ due to the planet. $\Delta(t)$ is the magnification of the
source star with respect to time, minus the point-lens amplification
that would occur if the planet were not present. The x-axis shows time
from the maximum of the deviation in days. For each of the five
systems considered in our simulation, five deviations are shown,
chosen randomly from the set of events with a deviation of $\Delta >
0.01$ lasting at least 6 hours. Deviations include a mixture of close
approaches (for which $A_{max}$ is large) and distant approaches (for
which $A_{max}$ is small). This is because when $\alpha$ is small,
there are two general regions of deviation in the lens plane: one near
to the lens and one far from it, as shown in Figure 2. Deviations on
near approaches are due to crossing the region near the lens, and
deviations on far approaches are due to crossing the region far from
the lens. As $\alpha$ approaches 1 (i.e., for small $D_L$), these two
regions coalesce, and so for $D_L$ small, there are not two distinct
deviation types, but rather a continuum. For the smallest values of
$\alpha$ (i.e., the largest values of $D_L$, for a given lens mass),
deviations experienced in the outer region tend to be short-lived, so
that the longest deviations occur when the source track traverses the
central region.}
\end{figure}

\bibliography{Glissando}

\end{document}